\documentstyle{article}
\advance\voffset by -1 in
\advance\hoffset by -.5 in
\begin{document}
\thispagestyle{plain}
\def\a{\alpha}
\def\b{\beta}
\def\g{\gamma}
\def\d{\delta}
\def\e{\epsilon}
\def\k{\kappa}
\def\l{\lambda}
\def\x{\xi}
\def\f{\phi}
\def\j{\psi}
\def\z{\zeta}
\def\p{\partial}
\def\o{\omega}
\def\O{\Omega}
\def\Om0{\Omega^0}
\def\O1{\Omega^1}
\def\cA{{\cal A}}
\def\cB{{\cal B}}
\def\tcA{\tilde{\cal A}}
\def\tcB{\tilde{\cal B}}
\def\tf{\tilde{f}}
\def\tg{\tilde{g}}
\def\th{\tilde{h}}
\def\rarr{\rightarrow}
\def\mp{\mapsto}
\def\hL{\hat{L}}
\def\res{{\rm res}\:}
\def\tr{{\rm tr}\:}
\def\cF{{\cal F}}
\def\et{\eta}
\def\cV{{\cal V}}
\def\>{\rangle}
\def\<{\langle}
\def\Cal{\cal}
\def\({\left(}
\def\){\right)}
\hfill treves2.tex \today

\vspace{.2in}
\noindent \Large {\bf On Treves' Algebraic Characterization
of the KdV Hierarchy.}\normalsize

\vspace{.2in}
\centerline{{\bf L. A. Dickey}\footnote{Math. Dept., University of Oklahoma,
e-mail: ldickey@math.ou.edu}}

\vspace{.2in}
\begin{abstract} We have found a possibility to streamline the proof of the
Treves' theorem [1] on an algebraic characterization of the KdV hierarchy
which makes it significantly shorter, following essentially the logic of the
original proof.
\end{abstract}

\vspace{.2in}
{\bf 1.} One of possible methods to construct the
equations of the KdV hierarchy is based on the recursion formula
$$R'''_{n-1}+4uR'_{n-1}+2u'R_{n-1}=-4R'_n,~R_0=1. \eqno{(1.1)}$$
Quantities $R_n$ are differential polynomials in $u$, i.e., polynomials
in $u$ and its derivatives. There is a grading in the algebra of
differential polynomials of $u$: $w(u)=2$, $ w(\p)=1$ where
$\p=d/dx$. Eq. (1.1) determines $R_n$ up to a constant of integration.
Two sequences of $R_n$ obtained with different choice of constants
are related by a linear triangular transformation, i.e., terms of one
sequence are linear combinations of terms of the other sequence with
less or equal numbers.
 If the requirement is imposed that $\{R_n[u]\}$ are
homogeneous in weight differential polynomials, $w(R_n)=2n$, then
the recursion formula (1) uniquely determines all $R_n$'s. The
equations of the KdV hierarchy are $$\p_{t_n}u=R'_n[u],\eqno{(1.2)}$$
it is supposed that $u$ depends on parameters $t_n$. All $R_m$'s
are first integrals of each Eq. (1.2) in the sense that
$$\p_{t_m}R_n=\p\:Q_{nm}$$ where $Q_{nm}$ is a differential
polynomial (see, e.g., [2], [3]). Any linear combination of $R_n$ is
also a first integral. Two first integrals are equivalent if they
differ by a differential polynomial which is the derivative of another
polynomial.

Treves [1] gave the following criterion of the fact that a given differential
polynomial $P[u]$ is equivalent (differs by an exact derivative) to a
linear combination of $R_n$:\\

{\bf Theorem} (Treves). A
differential polynomial $P[u]$ is $\sum_nc_nR_n[u]+\p Q[u]$ where $Q$ is a
differential polynomial if and only if the following criterion is
satisfied: let the formal series $$-{2\over x^2}+a_0+\sum_2^\infty
a_kx^k,~a_k={\rm const}$$ be substituted for $u$ in $P[u]$. Then
$${\rm res}_xP\left[-{2\over x^2}+a_0+\sum_2^\infty a_kx^k\right]=0.
\eqno{(1.3)}$$ The residue ${\rm res}_x$ symbolizes, as usual, the
coefficient of $x^{-1}$.\\

The theorem is remarkable for the following two reasons. (1) This
criterion has a touch of that enigmatical universality which distinguishes
the celebrated Sato bilinear identity: there is no hint on the KdV equations
in Eq. (1.3), it has a very general character, and nevertheless the hierarchy
is invisibly present there. (2) It suggests the use of the ``trial functions'' in
the form of formal Laurent series in x in the study of differential polynomials.
This can work for other problems, too. For example, it is possible to prove
the following necessary and sufficient criterion for a differential polynomial of
$u$ to be a derivative of another differential polynomial: if an arbitrary
semi-infinite Laurent series $\sum_{-N}^\infty a_nx^n$ is substituted for $u$, the
residue of the obtained Laurent series is zero. The Treves theorem states
that if the class of trial series is restricted in an appropriate way then
the Hamiltonians of the KdV hierarchy should be added to the exact derivatives.\\

{\bf 2. Necessity of the criterion (1.3).}

{\it Proof.} With the second term, $\p Q[u]$, it is clear: when
any Laurent series of $x$ is substituted for $u$, then $Q[u(x)]$
is a Laurent series and its derivative, $\p Q[u(x)]$, has the
zero residue. Now we have to prove that the formal series in
powers of $x$
$$R_n\left[-{2\over x^2}+a_0+\sum_2^\infty a_kx^k\right]=\hat R_n(x)$$
satisfy (1.3). The sequence of formal series in powers of $x$:
$\{\hat R_n(x)\}$ satisfies the recursion formula
$$ \hat R'''_{n-1}(x)+4\left(-{2\over x^2}+a_0+\sum_2^\infty a_kx^k\right)
\hat R'_{n-1}(x)+2\left(-{2\over x^2}+a_0+\sum_2^\infty
a_kx^k\right)'\hat R_{n-1}(x) =-4\hat R'_n(x).$$ Taking $\hat
R_0=1$, the other $\hat R_n$ can be recovered from the sequence of the recursion
relations uniquely up to arbitrary
additive constants. The choice of the constants is irrelevant since if the
theorem is proven for one choice of constants it is true for all the others
since the residue is a linear functional. For
simplicity, the free of $x$ terms can be taken as zero.

Let us find $\hat R_1$ and $\hat R_2$. We have
$2(4x^{-3}+2a_2x+3a_3x^2+...)=-4\hat R'_1$ whence $$\hat
R_1={1\over x^2}- {1\over 2}(a_2x^2+a_3x^3+...).$$ Denote
$a_0+a_2x^2+a_3x^3+...=\phi$ and $T_1=\sum_0^\infty
T_{1,k}x^k=-(1/2)(a_2x^2+a_3x^3+...)$, so, $\hat R_1 =x^{-2}+T_1$.
The next recursion formula is
$$((-2)(-3)(-4)x^{-5}+T_1''')+4(-2x^{-2}+\phi)(-2x^{-3}+
T'_1)+2(4x^{-3}+\phi')(x^{-2}+T_1)=-4\hat R'_2.$$ The terms with
$x^{-5}$ go. The terms with $x^{-3}$ are $ -8a_0+8T_{1,0}=-8a_0$.
The terms with $x^{-2}$ are $-8T_{1,1}+8T_{1,1}=0$, and the terms
with $x^{-1}$ are $-16T_{1,2}-
8a_2+8T_{1,2}+4a_2=-8T_{1,2}-4a_2=4a_2-4a_2=0$. Now, $-4\hat
R_2'=(-8a_0) x^{-3}+$ (terms with nonnegative powers of $x$).
Therefore, $\hat R_2=A_2x^{- 2}+T_2$ where $A_2$ is a constant and
$T_2$ is a series in positive powers of $x$.

Now we can make a hypothesis that all $\hat R_m$ have a form $\hat
R_m=A_{m}x^{- 2}+T_m$ where $A_m$ is a constant and $T_m$ a series
in positive powers of $x$ $(T_{m-1,0}=0)$, and prove it by
induction. The recursion formula has the form:
$$(-2)(-3)(-4)A_{m-1}x^{-5}+T_{m-1}'''+4(-2x^{-2}+\phi)
(-2A_{m-1}x^{-3}+
T'_{m-1})$$$$+2(4x^{-3}+\phi')(A_{m-1}x^{-2}+T_{m-1})=-4R'_m.$$
Collect the terms with the same power of $x$:
\begin{eqnarray*}
x^{-5}:&&\quad -24A_{m-1}+16A_{m- 1}+8A_{m-1}=0,\\
x^{-3}:&&\quad -A_{m-1}a_0=A_m,\\
x^{-2}:&&\quad -8T_{m-1,1}+8T_{m-1,1}=0,\\
x^{-1}:&&\quad -16T_{m-1,2}-
8A_{m-1}a_2+8T_{m-1,2}+4A_{m-1}a_2=-8T_{m-1,2}-4A_{m-1}a_2.
\end{eqnarray*} In
the cases of $\hat R_1$ and $\hat R_2$ this term was zero. Therefore,
we make another hypothesis that it is always zero, and also prove it by
induction. Thus, we suppose
$$2T_{m-1,2}+A_{m-1}a_2=0. \eqno{(2.1)}$$ We continue:
\begin{eqnarray*}
x^0:&&\quad 6T_{m-1,3}-8\cdot 3T_{m-1,3}-8A_{m-1}a_3+4a_0T_{m-1,1}\\
&&\quad +2A_{m-1}\cdot 3a_3+8T_{m-1,3}=-4T_{m,1},\end{eqnarray*}
 or \begin{eqnarray*} x^0:&&\quad
-10T_{m-1,3}+4a_0T_{m-1,1}-14A_{m-1}a_3=-4T_{m,1},\\
x^1:&&\quad 4\cdot 3\cdot 2T_{m-1,4}-8\cdot 4T_{m-1,4}-8A_{m-1}a_4\\
&&\quad +4\cdot 2a_0T_{m-1,2}+8T_{m-1,4}+2A_{m-1}\cdot
4a_4=-8T_{m,2}
\end{eqnarray*} or,
taking into account (2.1), $ a_0a_2A_{m-1}=2T_{m,2}$. We had the
equation $-A_{m-1}a_0=A_m$. Therefore, $$x^{1}:\quad
A_ma_2+2T_{m,2}=0$$ which is nothing but our hypothesis (2.1) for
the next number $m$. The rest of equations determine $T_{m,3},...$ Now both the
hypotheses are proven and we
have
$$\hat R_m=A_mx^{-2}+T_m,~(T_m=T_{m,1}x+T_{m,2}x^2+...).$$

Take the residue: $${\rm res}_x\hat R_m=0$$ q.e.d.\\

{\bf 3. Sufficiency of the criterion (1.3). Beginning of the proof.}

A differential polynomial $P[u]$ is a polynomial $P(\xi_0,\xi_1,\xi_2,...)$
where $u,u',u'',...$ are substituted for $\xi_0,\xi_1,\xi_2,...$.
As it was said before, there is a grading: $w(u^{(n)})=n+2$, $w(\p)=1$. If all terms
of a polynomial $P$ have the same weight $\k$ then $P(\l^2u,\l^3u',\l^4u''
...)=\l^\k P(u,u',u'',...)$.

We must prove the second part of the Treves theorem:\\

 A  differential polynomial $P[u]$
satisfying the Treves condition $$\res_x\:P\left[-2/x^2+a_0+\sum_2^\infty
a_n{x^n\over n!}\right]=0\eqno{(3.1)}$$ for any $\{a_n\}$ is a sum of an exact
derivative $\p S[u]$ and a linear combination of the KdV polynomial $R_k[u]$.\\

{\it Proof.} First we are proving a lemma:\\

{\bf Lemma.} If a differential polynomial satisfies the Treves condition (3.1)
then so does each homogeneous in weight component of this polynomial.\\

{\em Proof of the lemma.} Let $P=\sum P_\k$ where $P_\k$ a homogeneous polynomial
of weight $\k$. Since $\{a_n\}$ are arbitrary, we can replace them by $a_n\l^
{n+2}$. Now,
$$\res_x\:\sum P_\k\left(-2/x^2+\l^2a_0+\sum_2^\infty \l^{n+2}a_n{x^n\over n!},
4/x^3+\sum_2^\infty \l^{n+2}a_n{x^{n-1}\over {n-1}!},-12/x^4+
\sum_2^\infty \l^{n+2}a_n{x^{n-2}\over (n-2)!},...
\right)$$$$=
\res_x\:\sum l^\k P_\k\left(-2/(lx)^2+a_0+\sum_2^\infty \a_n{(lx)^n\over n!},
4/(lx)^3+\sum_2^\infty a_n{(lx)^{n-1}\over {n-1}!},-12/(lx)^4+
\sum_2^\infty a_n{(lx)^{n-2}\over (n-2)!},...
\right)$$$$=\sum\l^{\k-1}\res_xP_\k\left[-2/x^2+a_0
+\sum_2^\infty a_n{x^n\over n!}\right].$$ If this is zero, then each term is
zero since $\l$ is arbitrary, q.e.d.

Therefore, we can consider each component of weight $\k$ separately.
It is easy to prove by induction, using the recursion relation
for $R_k$ (1.1), that all $R_k$ (recall that $w(R_k)=2k$)
contain the term $u^k$ with
a non-zero coefficient. Thus, if the given polynomial $P_{\k}$, $\k=2k$
contains this term, then there is a constant $c$ such that $P-cR_k$ is without
this term and satisfies the Treves condition (3.1) since both $P_\k$
and $R_k$ do. If $\k$ is odd, $P_\k$ cannot have terms $cu^l$.
Further we are showing that a polynomial of the weight
$\k$ satisfying (3.1) and without terms $cu^l$ is an exact derivative.
Let $P$ be such a polynomial. Any non-zero polynomial can always by
transformed by adding an exact derivative to a ``reduced form'' which means
that all its monomials have a form $$c(u^{(j_1)})^{q_1}\cdots (u^{(j_\mu)})^
{q_\mu},~j_1<...<j_\mu,~q_\mu\geq 2.$$ The number $\mu$ is the order of the term.
After this reduction, the polynomial $P$ still satisfies (1.1) since an
exact derivative always does, and it does not have terms $cu^l$.
A reduced polynomial cannot be an exact
derivative. Therefore, what we need to prove is that $P$ is identically
zero. Making an obvious change of variables, we will write the Treves condition
as $$\res_x\:P\left[1/x^2+a_0+\sum_2^\infty a_n{x^n\over n!}\right]=0
$$ or
$$\res_xP\left(1/x^2+a_0+\sum_2^\infty a_n{x^n\over
n!},\:-2/x^3+\sum_2^\infty a_n{x^{n-1}\over(n-1)!},...\right)=0. \eqno{(3.2)}$$
 We can
differentiate this equation with respect to $a_0$. Since $a_0$ enters
only the first argument of $P(u,u',u'',\:...)$, this will be
$$\res_x {\p\over\p u}P\left(1/x^2+a_0+\sum_2^\infty a_n{x^n\over
n!},\:-2/x^3+\sum_2^\infty a_n{x^{n-1}\over(n-1)!},\:...\right)=0. $$
We can repeat this operation again and again until the polynomial does
not contain $u$ and still is not zero if it was not initially, since
there is no term $u^k$ and the others contain besides $u$ other
variables, $u',u''...$. Moreover, it preserves its reduced form. Thus, we can
assume that $P$ does not contain $u$ and
is $P(u',u'',...)$. The Treves condition is then
$$\res_xP\left(-2/x^3+\sum_2^\infty a_n{x^{n-1}\over(n-1)!},\:
6/x^4+\sum_2^\infty a_n{x^{n-2}\over(n-2)!},\:...\right)=0.$$
Now take the derivative with respect to $a_j$:
$$\res_x\sum_{i=1}^j{\p^i\over\p x^i}\left({x^j\over j!}\right){\p\over\p u^i}
 P\left(-2/x^3+\sum_2^\infty a_n{x^{n-1}\over(n-1)!},\:
6/x^4+\sum_2^\infty a_n{x^{n-2}\over(n-2)!},\:...\right)=0.$$
Adding an exact derivative, ``integrating by parts'', we do not change the residue:
$$\res_x{x^j\over j!}{\d\over\d u} P\left(-2/x^3+\sum_2^\infty a_n{x^{n-1}\over
(n-1)!},\:6/x^4+\sum_2^\infty a_n{x^{n-2}\over(n-2)!},\:...\right)=0,~j=2,
3,...$$ where $\d/\d u=\sum_0^\infty(-1)^i\p^i(\p/\p u^{(i)})$ ($\p=\p/\p x$)
is the variational derivative. Since $\p P/\p u=0$, $${\d P\over\d u}=-\p{\d P\over
\d u'}=\p\sum_0^\infty(-1)^{i+1}\p^i{\p P\over\p u^{(i+1)}}=\p Q,~w(Q)=\k-3.$$ Then
the last equation, being integrated by parts, takes the form
$$\res_x x^j Q\left(-2/x^3+\sum_2^\infty a_n{x^{n-1}\over
(n-1)!},\:6/x^4+\sum_2^\infty a_n{x^{n-2}\over(n-2)!},\:...\right)=0,~j=1,
2,...$$ which means that the negative (principal) part of the Laurent expansion of
the function $$x Q\left(-2/x^3+\sum_2^\infty a_n{x^{n-1}\over
(n-1)!},\:6/x^4+\sum_2^\infty a_n{x^{n-2}\over(n-2)!},\:...\right)\eqno{(3.3)}$$
vanishes.

If we prove that $Q=0$, it will follow that $\d P/\d u=0$, i. e., $P$ is
an exact derivative which is impossible for a non-zero reduced differential
polynomial. Thus, it will be proven that reduced $P$ is zero, or that $P$ is
an exact derivative.

The homogeneity means
$Q(\l^3\xi_1,\l^4\xi_2,...)=\l^{\k-3}Q(\xi_1,\xi_2,...)$. Taking the
derivative with respect to $\l$ and letting $\l=1$, we get the Euler
identity $$3\xi_1\p Q/\p\xi_1+4\xi_2\p Q/\p\xi_2+...=(2k-
3)Q.\eqno{(3.4)}$$ Using the fact that $Q$ is homogeneous, one can
rewrite (3.3) as
$${1\over x^{\k-4}} Q\left(-2+\sum_2^\infty a_n{x^{n+2}\over
(n-1)!},\:6+\sum_2^\infty a_n{x^{n+2}\over(n-
2)!},\:...\right).\eqno{(3.5)}$$

{\bf 4. The proof continued.}

Now, we must expand (3.5) in powers of $x$. If doing this
directly, the expression is too involved. The following trick (Treves) simplifies
the task. Let $$\eta_j(x)=\sum_{n=0}^\infty a_{j+n}{x^n\over n!}.$$
The conversion of this formula is $a_j=\sum_{n=o}^\infty (-
1)^n\eta_{j+n}x^n/ n!$. Indeed, $$\sum_{n=o}^\infty (-
1)^n\eta_{j+n}{x^n\over n!}=\sum_{n=0}^\infty(-1)^n\sum_{m=0}^\infty
a_{j+n+m}{x^m\over m!}{x^n\over n!}=\sum_{p=0}^\infty a_{j+p}{x^p
\over p!}\sum_{n=0}^p(-1)^n{p\choose n}=a_j.$$ The expression (3.5)
takes the form $${1\over x^{\k-4}}Q\left(-2+\eta_1x^3,\:6+\eta_2x^4,
\:...,(-1)^j(j+1)!+\eta_jx^{j+2},\:...\right).$$ Quantities $\eta_j$ are not
independent since $a_1=0$. This gives a relation $\sum_{n=0}^\infty
(-1)^n\eta_{1+n}(x^n/ n!)=0$ or
$$\eta_1=\sum_{n=1}^\infty (-1)^{(n+1)}\eta_{1+n}{x^n\over n!}$$
whence the expression (3.5) is $${1\over x^{\k-4}}Q\left(-2+\sum_{n=1}
^\infty (-1)^{(n+1)}\eta_{1+n}{x^{n+3}\over n!},
\:6+\eta_2x^4,\:...\right).\eqno{(4.1)}$$ Expand $Q(\cdot)$ in powers of
$\eta$'s. We have $${\p \over\p\eta_j}Q(\cdot)=x^{j+2}D_jQ(\cdot)~{\rm
where}~D_j={\p\over\p\xi_j}+{(-1)^j\over(j-1)!}{\p \over\p\xi_1}$$ Therefore,
the expression (4.1) is
$$\sum_{q_2,...,q_r}{D_2^{q_2}\cdots D_r^{q_r}Q(\theta)\over q_2!\cdots q_r!
x^{\k-4-\l(q)}}\eta_2^{q_2}\cdots\eta_r^{q_r}\eqno{(4.2)}$$ where $$
(\theta)=(-2!,3!,...,(-1)^\mu(\mu+1)!),~  \l(q)=
\sum_{j=2}^r(j+2)q_j.$$

We had independent variables $a_2,a_3,...,x$ Now we changed them to new
independent variables $\eta_2,\eta_3,...,x$. The condition that the expression
(4.2) does not contain negative powers of $x$ becomes
$$D_2^{q_2}\cdots D_r^{q_r}Q(\theta)=0~{\rm when}~\l(q)<
\k-4.\eqno{(4.3)}$$

It is convenient to replace the operators $D_j$ by $L_j$:$$ L_2=D_2~{\rm and}
~L_j=D_j+(j-1)^{-1}D_{j-1},~j>2.$$ We have, $L_j=\p/\p\xi_j+\p/\p\xi_{j-
1}$. This is a triangular change of operators: $L_j$ is a
linear combination of $D_i$ with $i\leq j$ and vice versa. Therefore,
we can replace (4.3) by an equivalent equation
$$L_2^{q_2}\cdots L_r^{q_r}Q(\xi_1,\xi_2,...)|_{\xi_1=-
2,\xi_2=6,...}=0,~{\rm when}~\l(q)<\k-4.\eqno{(4.4)}$$

{\bf 5. End of the proof.}\\

{\bf Lemma.} The condition (4.3) (or, equivalently, (4.4)), implies
$Q\equiv 0$.\\

{\it Proof.} We use induction on $\mu$, the number of arguments of the polynomial
$Q(\xi_1,...,\xi_\mu)$. For $\mu=1$ the theorem is trivial. Indeed, since $Q$ is a
homogeneous polynomial of weight $\k-3$, and $\xi_1$ has the weight 3, $Q=c\xi_1
^{(\k-3)/3}$ if $\k-3$ is divisible by 3 and zero otherwise. If $c\neq 0$, then
$Q(-2)\neq 0$ which contradicts Eq. (4.3). Thus, $c=0$ and $Q=0$. Let the lemma
be proven for $\mu-1$.

We have $Q=\sum_{r_1,...,r_\mu}a_{r_1,...,r_\mu}\xi_1^{\mu_1}\cdots\xi_\mu^{r_\mu}$ where
$$3r_1+4r_2+...+(\mu+2)r_\mu=\k-3.\eqno{(5.1)}$$ This homogeneity
implies the Euler equation $$(3\xi_1\p_{\xi_1}+4\xi_2\p_{\xi_2}
+...+(\mu+2)\xi_\mu\p_{\xi_\mu})Q(\xi)=(\k-3)Q(\xi).\eqno{(5.2)}$$
Besides,  $$(\p_{\xi_1}+\p_{\xi_2})^{q_2}
(\p_{\xi_2}+\p_{\xi_3})^{q_3}\cdots(\p_{\xi_{\mu-
1}}+\p_{\xi_\mu})^{q_\mu}Q((\theta))
=0\eqno{(5.3)}$$ $${\rm when}~\l(q)=4q_2+5q_3+\cdots+(\mu+2)q_\mu<\k-
4.$$ One has to prove that under these conditions $Q\equiv 0$.

Let us apply the operator $\p_{\xi_1}^{i_1}\cdots\p_{\xi_{\mu-1}}^{
i_{\mu-1}}\p_{\xi_\mu}^{q-i_1-...-i_{\mu-1}-1}$ to (5.2) and put
$(\xi)=(\theta)$:
$$(-3!\:\p_{\xi_1}+4!\:\p_{\xi_2}+\cdots+(-1)^{\mu-1}(\mu+1)!\:\p_{\xi_{\mu-1}})
\p_{\xi_1}^{i_1}\cdots\p_{\xi_{\mu-1}}^{
i_{\mu-1}}\p_{\xi_\mu}^{q-i_1-...-i_{\mu-1}-1}Q((\theta))$$$$+
(-1)^\mu(\mu+2)!\:\p_{\xi_1}^{i_1}\cdots\p_{\xi_{\mu-1}}^{
i_{\mu-1}}\p_{\xi_\mu}^{q-i_1-...-i_{\mu-1}}Q((\theta))-A
$$ where $$A=c\p_{\xi_1}^{i_1}\cdots\p_{\xi_{\mu-1}}^{
i_{\mu-1}}\p_{\xi_\mu}^{q-i_1-...-i_{\mu-1}-1}Q((\theta))$$

Suppose it is already proven that all partial derivatives of order
$q-1$ vanish. Then we have $A=0$ and
$$\p_{\xi_1}^{i_1}\cdots\p_{\xi_{\mu-1}}^{
i_{\mu-1}}\p_{\xi_\mu}^{q-i_1-...-i_{\mu-1}}Q((\theta))$$$$=
\((-1)^\mu{3!\over(\mu+2)!}\p_{\xi_1}+(-1)^{\mu+1}{4!\over(\mu+2)!}
\p_{\xi_2}+\cdots+{(\mu+1)!\over(\mu+2)!}\p_{\xi_{\mu-
1}}\)\times$$$$\times
\p_{\xi_1}^{i_1}\cdots\p_{\xi_{\mu-1}}^{
i_{\mu-1}}\p_{\xi_\mu}^{q-i_1-...-i_{\mu-1}-1}Q((\theta))$$
We have diminished the order of the derivative with respect to
$\xi_{\mu}$ by 1. We can proceed until the derivative disappears at all:
$$\p_{\xi_1}^{i_1}\cdots\p_{\xi_{\mu-1}}^{
i_{\mu-1}}\p_{\xi_\mu}^{q-i_1-...-i_{\mu-1}}Q((\theta))$$$$=
\((-1)^\mu{3!\over(\mu+2)!}\p_{\xi_1}+(-1)^{\mu+1}{4!\over(\mu+2)!}
\p_{\xi_2}+\cdots+{(\mu+1)!\over(\mu+2)!}\p_{\xi_{\mu-
1}}\)^{q-i_1-...-i_{\mu-1}}\times$$$$\times
\p_{\xi_1}^{i_1}\cdots\p_{\xi_{\mu-1}}^{i_{\mu-1}}Q((\theta)).$$
Thus, computing a derivative of $q$th order one can always replace
$\p_{\xi_\mu}$ by the expression in parentheses in the last two
formulas. Denote this expression as $\p_{\xi_\mu}^*$. In
particular, (5.3) implies  $$(\p_{\xi_1}+\p_{\xi_2})^{q_2}
(\p_{\xi_2}+\p_{\xi_3})^{q_3}\cdots(\p_{\xi_{\mu-
1}}+\p_{\xi_\mu}^*)^{q_\mu}Q((\theta))=0\eqno{(5.4)}$$ if
$4q_2+5q_3+\cdots+(\mu+2)q_ \mu<\k-4$. Derivations
$\p_{\xi_1},...,\p_{\xi_{\mu-1}},\p_{\xi_\mu}^*$ can be expressed
as linear combinations of $(\p_{\xi_1}+\p_{\xi_2}),
...,(\p_{\xi_{\mu-1}}+\p_{\xi_\mu}^*)$, therefore any partial
derivative of order $q<(\k-4)/(\mu+2)$ of $Q$ at the point
$(\theta)$ is a linear combination of the expressions in the left-hand side of
(5.4), and, therefore, vanishes. The induction is proven.

This is still not the whole story. We have proven only that the
derivatives of order $<(\k-4)/(\mu+2)$ vanish at the point
$(\theta)$. In particular $\p_{\xi_\mu}^jQ(\theta)=0$ when $j<(\k-4)/(\mu+2)$.
On the other hand, this derivative vanishes when $j>(\k-3)/(\mu+2)$ since
$r_\mu\leq(\k-3)/(\mu+2)$ (see (5.1)). If $\k-3$ is not divisible by $\mu+2$,
this means that the derivatives of all orders with respect to $\xi_\mu$ vanish.
Q does not depend on $\xi_\mu$. Since we use the induction on $\mu$, we have
assumed that for $\mu-1$ the theorem is proven, so $Q\equiv 0$. If $(\k-3)/(\mu
+2)$ is an integer, say $n$, then the only non-zero derivative with respect to
$\xi_\mu$ at $(\theta)$ can be the $n$th one. Then, $Q$ has a form $Q^*\cdot(\xi_\mu
-(-1)^\mu(\mu+1)!)^n$. However, this is homogeneous only if $Q^*\equiv Q\equiv 0$ q.e.d.
The theorem is also proven.\\

{\bf References.}\\

[1] F. Treves, An algebraic characterization of the Korteweg-de Vries
hierarchy, Duke Math. Journal, 108, No. 2, 251-294, 2001\\

[2] I. M. Gelfand and L. A. Dickey, Asymptotic of the resolvent of
Sturm-Liouville's equations and algebra of the Korteweg-de Vries
equation, Uspehi Math. Nauk (Russian Math. Surveys), 30, No. 5,
87-100, 1975\\

[3] L. A. Dickey, Soliton equations and Hamiltonian systems,
Advanced Series in Math. Physics, World Scientific, 1st ed. vol 12,
1991, 2nd ed. vol 26, 2003

\end{document}